\documentclass[article,preprint,12pt,unsrt]{elsarticle}
\usepackage{a4wide}
\usepackage{natbib} 			
\usepackage{graphicx}
\usepackage{amsfonts}
\usepackage{amsmath}
\usepackage{booktabs}
\usepackage{epsfig}
\usepackage{stmaryrd}
\usepackage{hyperref}
\hypersetup{citecolor=black,linkcolor= black}

\begin{document}
\title{
Can the bivariate Hurst exponent be higher than an average of the separate Hurst exponents?}
\author{Ladislav Kristoufek}
\ead{kristouf@utia.cas.cz}
\address{Institute of Information Theory and Automation, Academy of Sciences of the Czech Republic, Pod Vodarenskou Vezi 4, 182 08, Prague 8, Czech Republic\\
Institute of Economic Studies, Faculty of Social Sciences, Charles University, Opletalova 26, 110 00, Prague 1, Czech Republic
}

\begin{abstract}
In this note, we investigate possible relationships between the bivariate Hurst exponent $H_{xy}$ and an average of the separate Hurst exponents $\frac{1}{2}(H_x+H_y)$. We show that two cases are well theoretically founded. These are the cases when $H_{xy}=\frac{1}{2}(H_x+H_y)$ and $H_{xy}<\frac{1}{2}(H_x+H_y)$. However, we show that the case of $H_{xy}>\frac{1}{2}(H_x+H_y)$ is not possible regardless of stationarity issues. Further discussion of the implications is provided as well together with a note on the finite sample effect.
\end{abstract}


\journal{Physica A}

\maketitle


\newpage

\section{Introduction}

Generalization of power-law correlations (long-term memory, long-range dependence) into a bivariate setting has brought a wide range of possibilities for studying connections between various series. These power-law cross-correlations have become popular especially in econophysics with applications to numerous financial series \cite{Podobnik2009,Podobnik2009a,He2011,Cao2012,Lin2012,Wang2013b}. Formally, the bivariate long-term memory translates into a power-law decay of the cross-correlation function $\rho_{xy}(k)$ with lag $k$ so that $\rho_{xy}(k) \propto k^{2H_{xy}-2}$ for $k \rightarrow +\infty$. The cross-correlation function is thus hyperbolically decaying in the same manner as the auto-correlation function in the univariate case. Alternatively, the bivariate long-range dependence can be defined in the frequency domain via a divergence of spectrum close to the origin. Specifically, the cross-spectrum $f_{xy}(\omega)$ with frequency $\omega$ has a form of $|f_{xy}(\omega)|\propto \omega^{1-2H_{xy}}$ for $\omega \rightarrow 0+$. The bivariate Hurst exponent $H_{xy}$ measures a strength of such power-law cross-correlations\footnote{Alternatively, parameters $\alpha$, $\alpha_{xy}$ $\lambda_{xy}$ or $d_{12}$ are used as measures of power-law cross-correlations in the literature.} \cite{Robinson2008,Sela2012}.

The ideas of long-range cross-correlations have been reflected in an introduction of various estimators of the bivariate Hurst exponent. These are usually bivariate generalizations of the univariate estimators -- detrended cross-correlation analysis (DCCA or DXA) \cite{Podobnik2008,Zhou2008,Jiang2011}, height cross-correlation analysis (HXA) \cite{Kristoufek2011} and detrending moving-average cross-correlation analysis (DMCA) \cite{Arianos2009,He2011a}. In addition, new correlation coefficients have been proposed based on the ideas of the bivariate estimators. Most notably, Zebende \cite{Zebende2011,Zebende2013} introduces the DCCA-based correlation coefficient and Kristoufek \cite{Kristoufek2014a} adds the DMCA-based correlation coefficient. These two play an important role in our further discussion. Several tests of the power-law cross-correlations have been introduced as well \cite{Podobnik2011,Kristoufek2013a}.

In the applied literature, the main focus is usually put on the bivariate Hurst exponent $H_{xy}$ and its comparison to the Hurst exponents of the separate processes, $H_x$ and $H_y$. Numerically, it has been shown that various theoretical processes imply $H_{xy}=\frac{1}{2}(H_x+H_y)$ \cite{Sela2012,Podobnik2011,Kristoufek2013,Kristoufek2015}. Several processes having $H_{xy}<\frac{1}{2}(H_x+H_y)$ have been proposed as well \cite{Sela2012,Kristoufek2013}. However, various studies report that the bivariate Hurst exponent is higher than the average of the separate processes, i.e. $H_{xy}>\frac{1}{2}(H_x+H_y)$ \cite{He2011,Wang2013b,He2011b,Wang2013,Oswiecimka2014}. An unanswered question remains -- are all these three possibilities feasible? More specifically, it is only not obvious whether the last option is feasible as the former two have been shown to exist analytically. In this short paper, we answer the posed question. The next section provides the needed instruments. The last section brings some novel insights into the topic with a discussion of implications.


\section{Methodology}

For studying the relationship between the bivariate Hurst exponent $H_{xy}$ and the separate Hurst exponents $H_x$ and $H_y$, we recall several concepts from both time and frequency domains. We present the spectrum coherence (frequency domain) and the DCCA and DMCA correlation coefficients (time domain). Both concepts are essential here.

The squared spectrum coherency is defined for two stationary series $\{x_t\}$ and $\{y_t\}$ with existing spectra $f_{xy}(\omega)$, $f_{x}(\omega)$ and $f_{y}(\omega)$ at frequency $0\le \omega \le \pi$. Squared spectrum coherency $K_{xy}^2(\omega)$ is defined as 
\begin{equation}
K_{xy}^2(\omega)=\frac{|f_{xy}(\omega)|^2}{f_{x}(\omega)f_{y}(\omega)}
\end{equation}
for a given frequency $\omega$. The squared coherence can be understood as a squared correlation between processes $\{x_t\}$ and $\{y_t\}$ at frequency $\omega$. Note that it holds that $0\le K_{xy}^2(\omega)\le1$ for all $\omega$ \citep{Wei2006}.

The detrended cross-correlation coefficient $\rho_{DCCA}(s)$ for scale $s$ \cite{Zebende2011} combines the detrended fluctuation analysis (DFA) \citep{Peng1993,Peng1994,Kantelhardt2002} and the detrended cross-correlation analysis (DCCA) \citep{Podobnik2008,Zhou2008,Jiang2011}. The DCCA-based coefficient for scale $s$ is defined as
\begin{equation}
\rho_{DCCA}(s)=\frac{F^2_{DCCA}(s)}{F_{DFA,x}(s)F_{DFA,y}(s)},
\label{rho}
\end{equation}
where $F^2_{DCCA}(s)$ is a detrended covariance between profiles of series $\{x_t\}$ and $\{y_t\}$ based on a window of size $s$, and $F^2_{DFA,x}$ and $F^2_{DFA,y}$ are detrended variances of profiles of the separate series, respectively, for a window size $s$. The detrending moving-average cross-correlation coefficient $\rho_{DMCA}(\lambda)$ for window size $\lambda$ \cite{Kristoufek2014a} connects the detrending moving average (DMA) procedure \citep{Vandewalle1998,Alessio2002} and the detrending moving-average cross-correlation analysis (DMCA) \citep{Arianos2009,He2011a}. The coefficient is defined as
\begin{equation} 
\rho_{DMCA}(\lambda)=\frac{F_{DMCA}^2(\lambda)}{F_{x,DMA}(\lambda)F_{y,DMA}(\lambda)},
\end{equation}
where $F^2_{DMCA}(\lambda)$, $F^2_{DMA,x}(\lambda)$ and $F^2_{DMA,y}(\lambda)$ are a detrended covariance between profiles of the examined series and detrended variances of the separate series, respectively, with a moving average parameter $\lambda$. Both coefficients have been shown to range between $-1 \le \rho_{DCCA}(s),\rho_{DMCA}(\lambda)\le 1$ analytically for all scales $s$ or windows sizes $\lambda$ \cite{Kristoufek2014a,Podobnik2011}.

\section{Discussion}

The squared spectrum coherency gives straightforward implications for the bivariate Hurst exponents. Rewriting the coherency using the definition of the power-law cross-correlations in the frequency domain, we obtain

\begin{equation}
\label{eq:coherence}
K_{xy}^2(\omega)=\frac{|f_{xy}(\omega)|^2}{f_{x}(\omega)f_{y}(\omega)}\propto \frac{\omega^{2(1-2H_{xy})}}{\omega^{1-2H_x}\omega^{1-2H_y}}=\omega^{2(H_x+H_y-2H_{xy})}.
\end{equation}
As the squared coherency lies between 0 and 1 for all frequencies, it does so for the long-range cross-correlations case of $\omega \rightarrow 0+$ as well. Therefore, this gives us two feasible and one infeasible possibilities:
\begin{itemize}
\item $H_{xy}=\frac{1}{2}(H_x+H_y) \Rightarrow 2(H_x+H_y-2H_{xy})=0 \Rightarrow \lim_{\omega\rightarrow 0+}{K_{xy}^2(\omega)\propto const.}$
\item $H_{xy}<\frac{1}{2}(H_x+H_y) \Rightarrow 2(H_x+H_y-2H_{xy})>0 \Rightarrow \lim_{\omega\rightarrow 0+}{K_{xy}^2(\omega)=0}$
\item $H_{xy}>\frac{1}{2}(H_x+H_y) \Rightarrow 2(H_x+H_y-2H_{xy})<0 \Rightarrow \lim_{\omega\rightarrow 0+}{K_{xy}^2(\omega)=+\infty} \Rightarrow \lightning $
\end{itemize}
This implies that for stationary processes, we cannot have $H_{xy}>\frac{1}{2}(H_x+H_y)$ as it is in contradiction with the bounded squared spectrum coherency. This also translates into the non-stationary case with pseudo-spectra. To make the claim for the non-stationary case stronger, we show the contradiction in the time domain as well.

Both the DCCA and DMCA coefficients are fixed between -1 and 1 for all feasible scales but also for both stationary and non-stationary specifications of the underlying processes \cite{Kristoufek2014a,Podobnik2011}. Similarly to the coherency case, we can rewrite the coefficients using the power-law correlations definition in the time domain. For this, we need to recall that for the long-range cross-correlated processes, we have $F^2_{DCCA}(s) \propto s^{2H_{xy}}$ for $s\rightarrow +\infty$ \cite{Podobnik2008} and $F^2_{DMCA}(\lambda) \propto \lambda^{2H_{xy}}$ for $\lambda \rightarrow +\infty$ \cite{Arianos2009} so that the correlation coefficients can be rewritten as
\begin{gather}
\rho_{DCCA}(s)=\frac{F^2_{DCCA}(s)}{F_{DFA,x}(s)F_{DFA,y}(s)} \propto \frac{s^{2H_{xy}}}{s^{H_x+H_y}} = s^{2H_{xy}-(H_x+H_y)} \nonumber \\
\rho_{DMCA}(s)=\frac{F^2_{DMCA}(\lambda)}{F_{DMA,x}(\lambda)F_{DMA,y}(\lambda)} \propto \frac{\lambda^{2H_{xy}}}{\lambda^{H_x+H_y}} = \lambda^{2H_{xy}-(H_x+H_y)}
\label{rho}
\end{gather}
We then have the same implications as for the frequency domain argument -- two feasible and one infeasible\footnote{Only the implications for DCCA are shown as the ones for DMCA are the same.}:
\begin{itemize}
\item $H_{xy}=\frac{1}{2}(H_x+H_y) \Rightarrow 2H_xy-(H_x+H_y)=0 \Rightarrow \lim_{s \rightarrow +\infty}{\rho_{DCCA}(s)}\propto const.$
\item $H_{xy}<\frac{1}{2}(H_x+H_y) \Rightarrow 2H_xy-(H_x+H_y)<0 \Rightarrow \lim_{s \rightarrow +\infty}{\rho_{DCCA}(s)}=0$
\item $H_{xy}>\frac{1}{2}(H_x+H_y) \Rightarrow 2H_xy-(H_x+H_y)>0 \Rightarrow \lim_{s \rightarrow +\infty}{\rho_{DCCA}(s)}=\pm\infty \Rightarrow \lightning$
\end{itemize}
The implications are thus the same as in the frequency domain but here, they hold also for non-stationary series. Having $H_{xy}>\frac{1}{2}(H_x+H_y)$ is thus not feasible in the power-law cross-correlations setting. The consequences of the presented results and the logic of arguments are far reaching.

First, the bivariate Hurst exponent is not necessarily equal to the average of the separate Hurst exponents. Second, unless at least one of the series is long-range correlated with $H>0.5$, the processes cannot be power-law cross-correlated with $H_{xy}>0.5$. Long-term memory of one of the underlying processes is thus needed and necessary. The power-law cross-correlations thus do not emerge out of nowhere but these are rather a by-product of the persistent separate process(es). This is well in hand with analytical results about long-range cross-correlated processes \cite{Podobnik2011,Kristoufek2013,Kristoufek2015}. Third, the case of $H_{xy}=\frac{1}{2}(H_x+H_y)$ is a natural limiting case for various processes with the non-zero squared coherency. These are not limited to the quite well studied and documented correlated ARFIMA processes or the mixtures of autoregressive and long-range dependent processes \cite{Sela2012,Podobnik2011,Kristoufek2013,Kristoufek2015} but they encompass rich possibilities. Fourth, the case of $H_{xy}>\frac{1}{2}(H_x+H_y)$ cannot happen which means that results suggesting it does fall victims to inefficient estimators of the bivariate Hurst exponent or are due to the finite sample effect\footnote{As the reviewers have suggested, the finite sample effect can play an important role. We focus on the time domain implications here but similar outcomes can be shown for the frequency domain as well. Recall that $|\rho_{DCCA}(s)|\le1$ for all scales $s$ (again the same logics can be applied to the DMCA coefficient). The proportionality in Eq. \ref{rho} can be written as 
\begin{equation}
\rho_{DCCA}(s)=Ks^{2H_{xy}-(H_x+H_y)}
\label{finite1}
\end{equation}
where $K$ is a proportionality term. Without a loss of generality, we assume $K>0$ and we thus focus on $0<\rho_{DCCA}(s)\le1$ (for $K<0$, we can perform a symmetric examination of the problem, and the problem is not interesting for $K=0$). Taking logarithm of Eq. \ref{finite1}, we have
\begin{equation}
\log K + (2H_{xy}-H_x-H_y)\log s=\log \rho_{DCCA}(s)\le0
\label{finite2}
\end{equation}
which implies
\begin{equation}
(2H_{xy}-H_x-H_y) \le -\frac{\log K}{\log s}.
\label{finite3}
\end{equation}
In the limiting case of $s \rightarrow +\infty$, we simply have $\lim_{s\rightarrow +\infty}{-\frac{\log K}{\log s}}=0$. However, for a finite sample case, we observe that even though the term $\log s$ goes to infinity, the divergence is quite slow. Therefore, the $\log K$ term can play a role in the finite sample analysis as $K$ can be either $0<K<1$, or $K=1$ or $K>1$ (note that we still assume $K>0$ here). Labelling the finite sample bias as $\zeta$, we have $H_{xy}\le\frac{H_x+H_y}{2}+\zeta$, where $\zeta > 0$, $\zeta = 0$ and $\zeta <0$ for $0<K<1$, $K=1$ and $K>1$, respectively. We can thus have a case when $H_{xy}>\frac{H_x+H_y}{2}$ caused by a finite sample bias for $0<K<1$. Nevertheless, such possibility still remains unfeasible in the asymptotic case.}. Interpretations based on such results are then misleading. And fifth, the case of $H_{xy}<\frac{1}{2}(H_x+H_y)$ is feasible and potentially interesting. Sela \& Hurvich \cite{Sela2012} refer to such processes as the anti-cointegration as the separate processes are long-range correlated but pairwise uncorrelated in a long-term horizon (at low frequencies). This is in evident opposition to the (fractional) cointegration for which it holds that $K^2_{xy}(\lambda)=1$ as $\lambda \rightarrow 0+$. The authors propose to use $d_{\rho}=d_{12}-\frac{d_1+d_2}{2}$ where $d_{12}$, $d_1$ and $d_2$ are fractional integration parameters for the joint long-term memory and the separate long-term memories, respectively, as a measure of power-law coherency. As we mainly function with the Hurst exponent definitions, we can rewrite the measure as $H_{\rho}=H_{xy}-\frac{H_x+H_y}{2}=d_{12}+\frac{d_1+d_2}{2}=d_{\rho}$ so that these are equivalent. If it holds that $H_{xy}=\frac{1}{2}{(H_x+H_y)}$, we have $H_{\rho}=0$, and for the anti-cointegration case, we have $H_{\rho}<0$. The latter case is only sparsely investigated in the literature \cite{Sela2012,Kristoufek2013} and it thus provides a relatively open field for further research, both theoretical and applied. 

\section*{Acknowledgements}

The author would like to thank the anonymous referees for valuable comments and suggestions which helped to improve the paper significantly. 
The research leading to these results has received funding from the European Union's Seventh Framework Programme (FP7/2007-2013) under grant agreement No. FP7-SSH-612955 (FinMaP). Support from the Czech Science Foundation under project No. 14-11402P and the Grant Agency of the Charles University in Prague under project No. 1110213 is also gratefully acknowledged.

\newpage

\section*{References}
\bibliography{Bibliography}

\begin{thebibliography}{10}

\bibitem{Podobnik2009}
B.~Podobnik, I.~Grosse, D.~Horvatic, S.~Ilic, P.~Ch. Ivanov, and H.~E. Stanley.
\newblock Quantifying cross-correlations using local and global detrending
  approaches.
\newblock {\em European Physical Journal B}, 71:243--250, 2009.

\bibitem{Podobnik2009a}
B.~Podobnik, D.~Horvatic, A.~Petersen, and H.~E. Stanley.
\newblock Cross-correlations between volume change and price change.
\newblock {\em PNAS}, 106(52):22079--22084, 2009.

\bibitem{He2011}
L.-Y. He and S.-P. Chen.
\newblock {Nonlinear bivariate dependency of price–volume relationships in
  agricultural commodity futures markets: A perspective from Multifractal
  Detrended Cross-Correlation Analysis}.
\newblock {\em Physica A}, 390:297--308, 2011.

\bibitem{Cao2012}
G.~Cao, L.~Xu, and J.~Cao.
\newblock Multifractal detrended cross-correlations between the chinese
  exchange market and stock market.
\newblock {\em Physica A}, 391:4855--4866, 2012.

\bibitem{Lin2012}
A.~Lin, P.~Shang, and X.~Zhao.
\newblock The cross-correlations of stock markets based on {DCCA} and
  time-delay {DCCA}.
\newblock {\em Nonlinear Dynamics}, 67:425--435, 2012.

\bibitem{Wang2013b}
F.~Wang, G.-P. Liao, X.-Y. Zhou, and W.~Shi.
\newblock Multifractal detrended cross-correlation analysis of power markets.
\newblock {\em Nonlinear Dynamics}, 72:353--363, 2013.

\bibitem{Robinson2008}
P.~Robinson.
\newblock Multiple local {Whittle} estimation in stationary systems.
\newblock {\em Annals of Statistics}, 36:2508--2530, 2008.

\bibitem{Sela2012}
R.~Sela and C.~Hurvich.
\newblock The average periodogram estimator for a power law in coherency.
\newblock {\em Journal of Time Series Analysis}, 33:340--363, 2012.

\bibitem{Podobnik2008}
B.~Podobnik and H.E. Stanley.
\newblock Detrended cross-correlation analysis: A new method for analyzing two
  nonstationary time series.
\newblock {\em Physical Review Letters}, 100:084102, 2008.

\bibitem{Zhou2008}
W.-X. Zhou.
\newblock Multifractal detrended cross-correlation analysis for two
  nonstationary signals.
\newblock {\em Physical Review E}, 77:066211, 2008.

\bibitem{Jiang2011}
Z.-Q. Jiang and W.-X. Zhou.
\newblock Multifractal detrending moving average cross-correlation analysis.
\newblock {\em Physical Review E}, 84:016106, 2011.

\bibitem{Kristoufek2011}
L.~Kristoufek.
\newblock Multifractal height cross-correlation analysis: A new method for
  analyzing long-range cross-correlations.
\newblock {\em EPL}, 95:68001, 2011.

\bibitem{Arianos2009}
S.~Arianos and A.~Carbone.
\newblock Cross-correlation of long-range correlated series.
\newblock {\em Journal of Statistical Mechanics: Theory and Experiment},
  3:P03037, 2009.

\bibitem{He2011a}
L.-Y. He and S.-P. Chen.
\newblock A new approach to quantify power-law cross-correlation and its
  application to commodity markets.
\newblock {\em Physica A}, 390:3806--3814, 2011.

\bibitem{Zebende2011}
G.F. Zebende.
\newblock {DCCA} cross-correlation coefficient: Quantifying level of
  cross-correlation.
\newblock {\em Physica A}, 390:614--618, 2011.

\bibitem{Zebende2013}
G.F. Zebende, M.F. da~Silva, and A.~Machado~Filho.
\newblock Dcca cross-correlation coefficient differentiation: Theoretical and
  practical approaches.
\newblock {\em Physica A}, 392:1756--1761, 2013.

\bibitem{Kristoufek2014a}
L.~Kristoufek.
\newblock Detrending moving-average cross-correlation coefficient: Measuring
  cross-correlations between non-stationary series.
\newblock {\em Physica A}, 406:169--175, 2014.

\bibitem{Podobnik2011}
B.~Podobnik, Z.-Q. Jiang, W.-X. Zhou, and H.~E. Stanley.
\newblock Statistical tests for power-law cross-correlated processes.
\newblock {\em Physical Review E}, 84:066118, 2011.

\bibitem{Kristoufek2013a}
L.~Kristoufek.
\newblock Testing power-law cross-correlations: Rescaled covariance test.
\newblock {\em European Physical Journal B}, 86:art. 418, 2013.

\bibitem{Kristoufek2013}
L.~Kristoufek.
\newblock Mixed-correlated {ARFIMA} processes for power-law cross-correlations.
\newblock {\em Physica A}, 392:6484--6493, 2013.

\bibitem{Kristoufek2015}
L.~Kristoufek.
\newblock On the interplay between short- and long-term memory in the power-law
  cross-correlations setting.
\newblock {\em Physica A}, accepted, 2015.

\bibitem{He2011b}
L.-Y. He and S.-P. Chen.
\newblock Multifractal detrended cross-correlation analysis of agricultural
  futures markets.
\newblock {\em Chaos, Solitons and Fractals}, 44:355--361, 2011.

\bibitem{Wang2013}
G.-J. Wang and C.~Xie.
\newblock Cross-correlations between the {CSI} 300 spot and futures markets.
\newblock {\em Nonlinear Dynamics}, 73:1687--1696, 2013.

\bibitem{Oswiecimka2014}
P.~Oswiecimka, S.~Drozdz, M.~Forczek, S.~Jadach, and J.~Kwapien.
\newblock Detrended cross-correlation analysis consistently extended to
  multifractality.
\newblock {\em Physical Review E}, 89:023305, 2014.

\bibitem{Wei2006}
W.W.S. Wei.
\newblock {\em Time series analysis: univariate and multivariate methods}.
\newblock Pearson Education, 2006.

\bibitem{Peng1993}
C.~Peng, S.~Buldyrev, A.~Goldberger, S.~Havlin, M.~Simons, and H.~Stanley.
\newblock Finite-size effects on long-range correlations: Implications for
  analyzing {DNA} sequences.
\newblock {\em Physical Review E}, 47(5):3730--3733, 1993.

\bibitem{Peng1994}
C.~Peng, S.~Buldyrev, S.~Havlin, M.~Simons, H.~Stanley, and A~Goldberger.
\newblock Mosaic organization of {DNA} nucleotides.
\newblock {\em Physical Review E}, 49(2):1685--1689, 1994.

\bibitem{Kantelhardt2002}
J.~Kantelhardt, S.~Zschiegner, E.~Koscielny-Bunde, A.~Bunde, S.~Havlin, and
  E.~Stanley.
\newblock {Multifractal Detrended Fluctuation Analysis of Nonstationary Time
  Series}.
\newblock {\em Physica A}, 316(1-4):87--114, 2002.

\bibitem{Vandewalle1998}
N.~Vandewalle and M.~Ausloos.
\newblock Crossing of two mobile averages: A method for measuring the roughness
  exponent.
\newblock {\em Physical Review E}, 58:6832--6834, 1998.

\bibitem{Alessio2002}
E.~Alessio, A.~Carbone, G.~Castelli, and V.~Frappietro.
\newblock Second-order moving average and scaling of stochastic time series.
\newblock {\em European Physica Journal B}, 27:197--200, 2002.

\end{thebibliography}
\bibliographystyle{unsrt}

\end{document}